\documentclass[journal]{IEEEtran}
\usepackage{xcolor}
\definecolor{color2b}{RGB}{0,0,255} 

\usepackage{soul}
\usepackage{graphicx}  
\usepackage{caption}   
\usepackage{float}     
\usepackage{dblfloatfix}

\usepackage[table]{xcolor}
\usepackage{array}
\usepackage{tabularx}

\usepackage{cite}

\ifCLASSINFOpdf
\else
\fi
\usepackage{amsmath}
\begin{document}
%
\title{Observation of Higher-Order Hybrid Bright-Dark Soliton Complexes in an NPR Mode-Locked Fiber Laser}
%
%
%

\author{Subrata Manna,
        Amala Jose,
        and~Nithyanandan Kanagaraj,~\IEEEmembership{Member,~IEEE}
\thanks{S. Manna, A. Jose, and N. Kanagaraj are with the Department of Physics, Indian Institute of Technology Hyderabad, Telangana, 502285 India e-mail: nithyan@phy.iith.ac.in}}
\maketitle

\begin{abstract}
Hybrid bright-dark soliton complexes offer profound insights into multi-component nonlinear wave dynamics, yet their systematic generation via artificial saturable absorbers remains unexplored. Here, we experimentally demonstrate a comprehensive hierarchy of higher-order hybrid soliton architectures within a single nonlinear polarization rotation (NPR)-based erbium-doped fiber laser. Continuous tuning of intracavity polarization states and pump power enables reproducible access to soliton structures, with stable mode-locking confirmed at 8 MHz and a signal-to-noise ratio exceeding 72 dB. Cross-phase modulation (XPM)-mediated inter-component coupling and birefringence-induced polarization evolution collectively govern the observed soliton multiplicity, with NPR's inherent tunability proving decisive in accessing higher-order hybrid soliton regimes unexplored in prior studies. These findings expand the fundamental taxonomy of dissipative solitons and establish NPR-based fiber lasers as a reconfigurable platform for complex nonlinear wave engineering.
\end{abstract}

%
\IEEEpeerreviewmaketitle

\section{Introduction}

Optical solitons in passively mode-locked fiber lasers have long captivated the photonics community, owing to their fundamental significance in nonlinear wave physics and their far-reaching technological implications~\cite{Singh2024}. Sustained by the precise equilibrium between group velocity dispersion (GVD) and the Kerr nonlinearity, gain and loss,  these self-localized wave packets exhibit exceptional robustness against perturbations, rendering them a natural yet powerful platform for investigating complex nonlinear light-matter interactions in guided-wave  systems~\cite{GonzalezTudela2024}. Beyond their intrinsic theoretical importance within the 
framework of the nonlinear Schr\"{o}dinger equation (NLSE), optical solitons have been 
harnessed across a remarkably broad spectrum of photonic applications, encompassing 
ultrafast pulse generation~\cite{Mohamed2024}, high-capacity optical 
communications~\cite{Corcoran2020}, precision frequency metrology~\cite{Fortier2019}, 
biomedical imaging~\cite{Marchand2021}, optical frequency comb 
synthesis~\cite{Sun2023}, and all-optical signal processing~\cite{Anagha2022}. The rich 
and intricate interplay between cavity architecture, intracavity dispersion management, and 
nonlinear mode-locking mechanisms~\cite{Pu2024} provides powerful leverage over soliton 
formation and evolution, enabling the generation of not only conventional bright~\cite{Mollenauer1980}  and dark solitons~\cite{Emplit1987}, but also more exotic multi-component hybrid  states~\cite{Kivshar1998} and vector soliton complexes~\cite{Zhao2010}.

Bright solitons are characterized by sharply localized intensity peaks rising above a near-zero continuous-wave background, whereas dark solitons manifest as well-defined intensity dips embedded within a nonzero background field. The coexistence and mutual interaction of these structurally contrasting wave entities arise from the combined influence of group velocity dispersion, self-phase modulation~(SPM), cross-phase modulation~(XPM), polarization-dependent nonlinearities, and birefringence-induced phase shifts accumulated 
over each round trip. Through careful and systematic engineering of these intracavity parameters, fiber laser cavities can support a remarkably broad continuum of pulse configurations ranging from elementary bright or dark pulses to higher-order multi-component bound states with intricate temporal and spectral profiles. This intrinsic configurational versatility positions passively mode-locked fiber lasers as an exceptionally fertile and controllable platform for exploring the rich physics of multi-component soliton interactions 
and dissipative nonlinear wave dynamics.

Among the simplest hybrid forms, dark-bright~(DB)~\cite{Kivshar1998} and 
bright-dark~(BD)~\cite{Christodoulides1988} soliton pairs have attracted sustained and growing experimental interest. These two-component structures, stabilized through XPM-mediated inter-component coupling, offer enhanced resistance to perturbations compared to their single-component counterparts~\cite{Kivshar1998b}, and hold considerable 
promise for long-distance optical communications~\cite{Kaur2024}, where multi-level information encoding within hybrid soliton waveforms represents a compelling paradigm. Early experimental demonstrations in this direction include the work of Meng \textit{et al.}~\cite{Meng2011} and Ning \textit{et al.}~\cite{Ning2012}, who reported stable bright-dark soliton emission in erbium-doped and dispersion-managed fiber lasers, highlighting polarization locking and the role of cavity design in stabilizing such states. These results were subsequently extended by Zhang \textit{et al.}~\cite{Zhang2016} and Gao \textit{et al.}~\cite{Gao2014}, confirming the generation of both BD and DB solitons across diverse mode-locking schemes and nonlinear material platforms.

Building upon these results, progressively more complex multi-component hybrid 
soliton states have begun to emerge in the experimental literature. Ma \textit{et~al.}~(2020) documented multi-soliton strings comprising cascaded BD groupings in a 
WS$_2$-based mode-locked system operating under elevated pump powers~\cite{Ma2020}, 
suggesting that extended cavity lengths and high intracavity energies favor the stabilization of complex soliton sequences, although in a polarization-sensitive manner. The bright-dark-bright~(BDB) soliton, representing a fully coherent three-component entity, was experimentally demonstrated in an EDFL mode-locked by a microfiber-based MoS$_2$ saturable 
absorber~\cite{Wu2021}, substantiating that such triplet states constitute single 
coherent entities rather than loosely bound multi-pulse complexes. In parallel, continued advances in novel two-dimensional and layered material platforms have further broadened the experimentally accessible family of hybrid states. Zhao \textit{et~al.} demonstrated 
dark-bright-bright~(DBB) soliton generation employing a ferromagnetic insulator 
Cr$_2$Ge$_2$Te$_6$ saturable absorber in an erbium-doped fiber 
laser~\cite{Zhao2020}, while Wang \textit{et~al.}~(2022) successfully realized BDB solitons 
using a chalcogenide-based Pb$_3$Sn$_4$FeSb$_2$S$_{14}$ absorber in an all-fiber 
erbium-doped configuration~\cite{Wang2022}. Most recently, Shi \textit{et~al.}~(2024) reported a particularly rich set of hybrid soliton states such as dark-bright-dark~(DBD), DBB, and dark-dark-bright~(DDB) within a ring 
cavity mode-locked by a germanene-based absorber~\cite{Shi2024}. These experimental observations find strong theoretical underpinning in multicomponent coupled nonlinear Schr\"{o}dinger models, which rigorously predict the existence of such inelastically coupled multi-component states, provided that strict group-velocity matching, 
precise inter-component phase relations, and sufficient intracavity energy are simultaneously maintained~\cite{Radhakrishnan1997}.

Despite these considerable and sustained developments, the existing body of work reveals a 
significant and largely unaddressed fundamental gap. While elementary DB and BD soliton 
pairs are now well-established across multiple laser platforms, more intricate, higher-order 
configurations such as the dark-dark-bright-bright~(DDBB) and 
bright-dark-bright-dark-bright~(BDBDB) soliton complexes have no prior experimental 
precedent in \textit{any} reported laser system. More critically, all prior demonstrations of complex hybrid soliton states have relied exclusively on material-based saturable absorbers, 
whose inherently fixed and non-reconfigurable nonlinear responses impose fundamental constraints on intracavity tunability, severely restricting access to the full parameter space of hybrid soliton dynamics. To the best of our knowledge, no experimental investigation has yet explored the generation of complex BD soliton architectures using artificial saturable absorber techniques, including nonlinear polarization rotation~(NPR), nonlinear optical loop mirrors~(NOLM), or nonlinear amplifying loop mirrors~(NALM). Further, artificial mode-locking mechanisms inherently possess dynamic, continuously reconfigurable, and in-situ adjustable control over polarization evolution and nonlinear phase accumulation, which are absent in material-based counterparts.
 \begin{figure}[htb]
    \centering
    \includegraphics[width=0.45\textwidth]{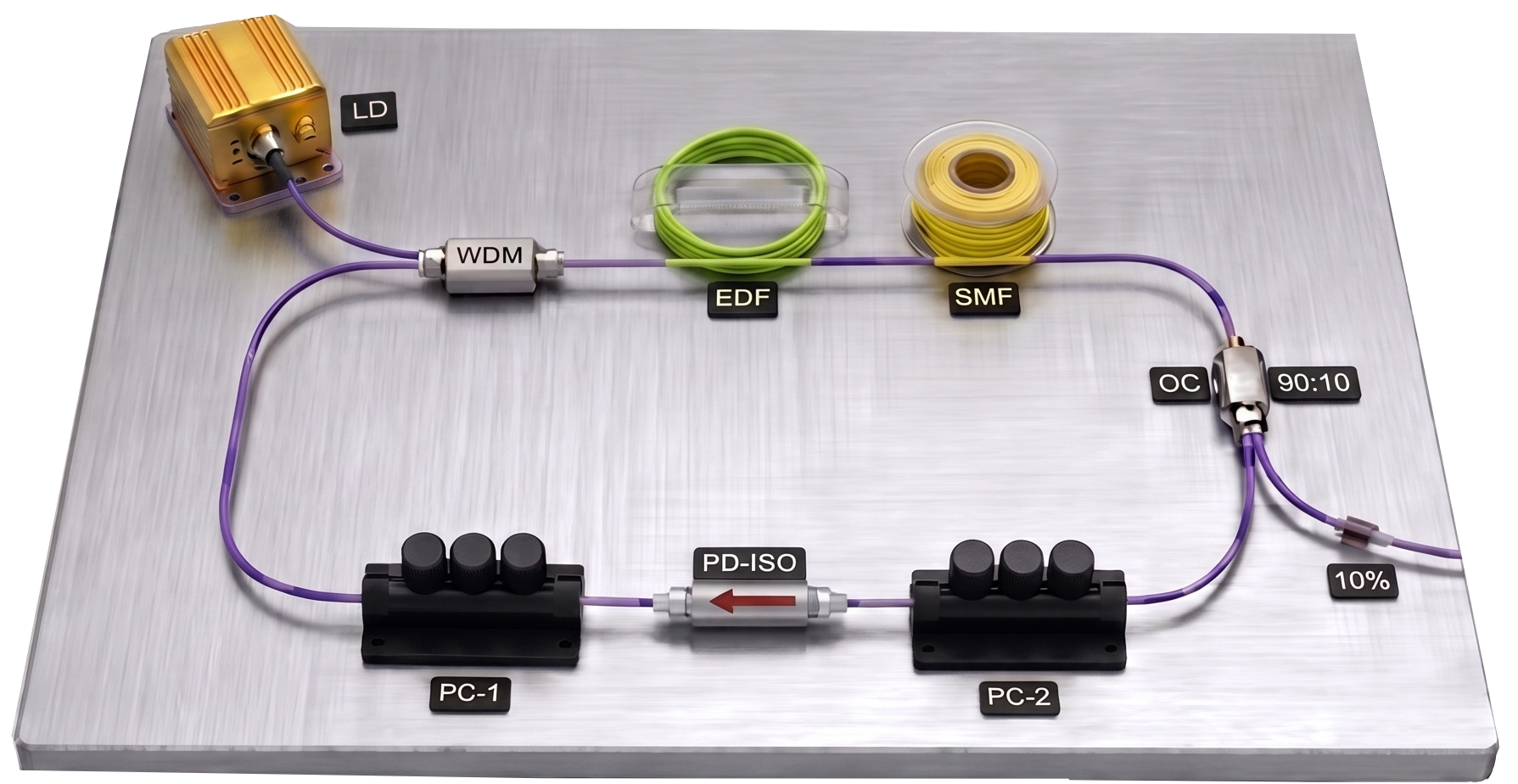} 
    \caption{Fig. 1. Schematic of the NPR-based mode-locked erbium-doped fiber laser cavity. EDF: erbium-doped fiber; WDM: wavelength division multiplexer; PC: polarization controller; PD-ISO: polarization-dependent isolator; OC: output coupler; SMF: single-mode fiber.}
    \label{1} 
\end{figure}
\begin{figure*}[htb]
    \centering
    \includegraphics[width=0.9\textwidth]{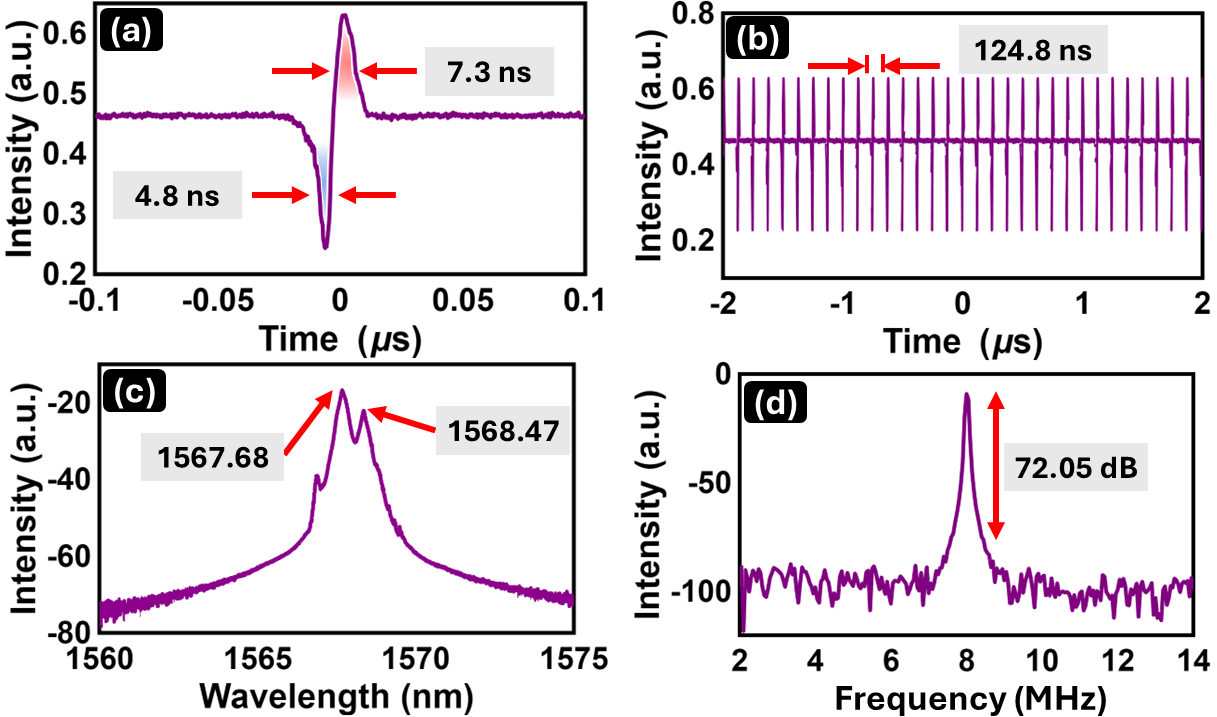} 
    \caption{Output characteristics of the DB soliton pair. (a)~Single-pulse profile with bright and dark component widths of 7.3~ns and 4.8~ns. (b)~Pulse train with 124.8~ns period at 8.015~MHz repetition rate. (c)~Dual-peaked optical spectrum at 1567.68~nm and 1568.47~nm. (d)~RF spectrum with SNR of 72.05~dB.
}
    \label{2} 
\end{figure*}
In this context, nonlinear polarization rotation~(NPR) emerges as a particularly versatile approach to artificial mode-locking. Unlike fixed-response material 
absorbers, NPR provides a dynamically and continuously tunable mode-locking mechanism, in which intracavity birefringence and nonlinear phase shifts can be precisely adjusted by rotating the polarization controllers, granting access to soliton regimes and multi-component configurations that remain experimentally uncharted in prior studies.  In this work, we utilize this property to demonstrate, to the best of our knowledge, the first experimental realization and systematic control of a comprehensive hierarchy of higher-order hybrid bright-dark soliton complexes—\textcolor{red}{including DB, BD, offset DB, DDB, BBD, BDD, DBD, DBB BDB, DDBB, and BDBDB}—within a single NPR-mode-locked erbium-doped fiber laser. The ability to generate, stabilize, and transition between these states highlights the versatility of NPR-based mode-locking and establishes a new benchmark for hybrid dissipative soliton dynamics in fiber laser systems.

\section{Experimental setup}
The schematic of the NPR-based mode-locked erbium-doped fiber laser~(EDFL) employed 
in this study is illustrated in Fig.~\ref{1}. The cavity is deliberately engineered 
to support a rich diversity of hybrid BD soliton structures through precise and 
continuous manipulation of intracavity polarization dynamics.
 
\begin{figure}[b]
    \centering
    \includegraphics[width=0.47\textwidth]{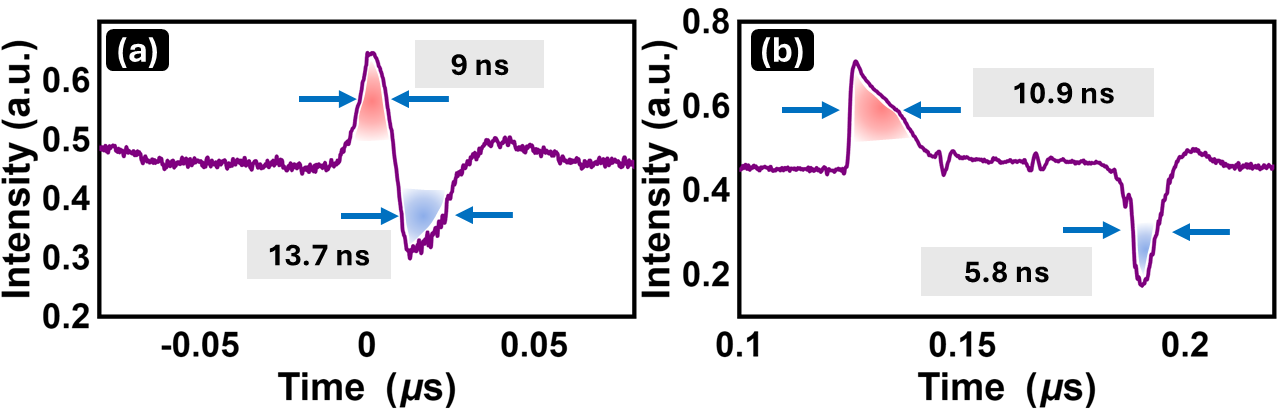} 
    \caption{Single-shot temporal profiles of (a) bright-dark and (b) offset bright-dark soliton pairs.}
    \label{3} 
\end{figure}
The active gain medium consists of a 3-meter erbium-doped fiber~(EDF), optically pumped 
by a 976~nm single-mode laser diode delivering a maximum output power of 1~W, 
corresponding to the $^{4}I_{15/2} \rightarrow\, ^{4}I_{11/2}$ absorption transition of 
the Er$^{3+}$ ion. The pump radiation is coupled into the EDF via a polarization-insensitive 
wavelength division multiplexer~(WDM), which simultaneously isolates residual pump 
light from the signal port.
The artificial saturable absorber effect is realized through NPR, implemented via two 
in-line polarization controllers~(PCs) and a polarization-dependent isolator~(PD-ISO). 
The PCs enable continuous adjustment of the intracavity polarization ellipse, while the 
PD-ISO enforces unidirectional pulse propagation and provides the intensity-dependent 
transmission essential for self-starting mode-locking. Together, these elements deliver reproducible control over nonlinear phase evolution, mode-locking threshold, and the dynamics of multi-component soliton formation. The output is extracted through the 10\% port of a 90:10 fiber output coupler~(OC). The total cavity length of 25.5~m, incorporating a 17-meter SMF-28 extension, yields a 
fundamental repetition rate of 8.015~MHz and a net anomalous group velocity dispersion of $\beta_{2} \approx -0.17$~ps$^{2}$ — essential for sustaining both bright and dark soliton constituents of the hybrid states investigated herein.

Output characterization is performed using a Yokogawa AQ6370D optical spectrum 
analyzer~(OSA, 0.2~nm resolution) for spectral analysis, a 4~GHz Tektronix MSO64 
oscilloscope paired with a 5~GHz Thorlabs DET08CFC/M InGaAs photodetector for 
temporal profiling, and an RF spectrum analyzer for mode-locking stability assessment via signal-to-noise ratio~(SNR) measurements. 
\section{Results and dicussion}
 The laser exhibits a well-defined pump power dependence, with CW lasing commencing at a threshold of 10~mW and self-starting mode-locking initiated upon increasing the pump power to 18~mW. The resulting bright soliton pulse train exhibits a uniform temporal period of 124.8~ns, in precise agreement with the calculated cavity round-trip time, corresponding to a fundamental repetition rate of 8.015~MHz — confirming stable, single-pulse-per-round-trip operation governed by the NPR-based dissipative soliton formation mechanism.
\begin{figure*}[htb]
    \centering
    \includegraphics[width=0.9\textwidth]{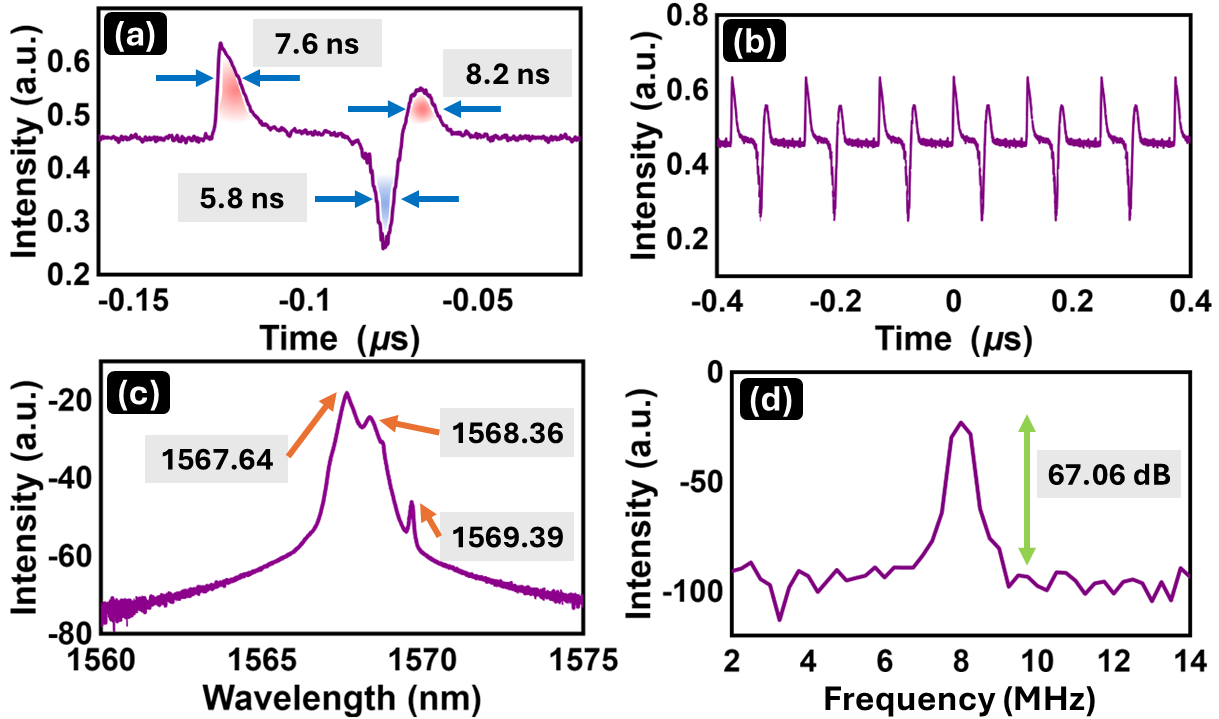} 
    \caption{Output characteristics of the BDB soliton. (a)~Single-pulse profile with component widths of 7.6~ns~(BP1), 5.8~ns~(DS), and 8.2~ns~(BP2). (b)~Pulse train at 8.015~MHz repetition rate. (c)~Optical spectrum with two spectral components at 1567.64 and 1568.36 nm. (d)~RF spectrum with SNR of 67.06~dB.}
    \label{bdb_f} 
\end{figure*}
\subsection{Dark-Bright~(DB) and Bright-Dark~(BD) Soliton Pairs}
At a pump power of 100~mW, systematic adjustment of the intracavity polarization state induced a spontaneous transition from single-component bright soliton operation 
to a two-component DB soliton pair, as shown in Fig.~\ref{2} (a) \& (b). 
Within the framework of coupled complex Ginzburg-Landau equations (CGLE),
the bright soliton — sustained by the balance of anomalous GVD and SPM in one polarization eigenstate — induces a periodic refractive index modulation via XPM that acts as an effective potential well for the orthogonally polarized field component. This XPM-mediated trapping stabilizes the dark soliton, which in turn exerts a reciprocal phase modulation on the bright component, establishing a self-consistent, mutually stabilized vector soliton state.
The single-pulse temporal profile~[Fig.~\ref{2}(a)] exhibits a pronounced asymmetric 
structure, with the bright and dark components displaying markedly different amplitudes 
and pulse widths. This asymmetry reflects the unequal nonlinear phase shifts accumulated 
by the two polarization components, arising from their differing peak intensities and the 
XPM cross-coupling coefficient $\sigma = 2/3$ governing orthogonally polarized fields 
in an isotropic fiber. The optical spectrum~[Fig.~\ref{2}(c)] displays a characteristic 
M-shaped dual-peaked profile with dominant emission at 1567.68~nm and 1568.47~nm, 
reflecting spectral interference between the two coupled polarization components. The observed asymmetric Kelly sidebands confirm anomalous dispersion operation and the soliton nature of the bright component. RF spectral analysis~[Fig.~\ref{2}(d)] yields an SNR exceeding 72~dB, confirming highly stable, low-noise mode-locked 
operation. The absence of an autocorrelation trace — consistent with prior 
reports~\cite{Emplit1987} — is attributable to the odd-symmetric field profile of the dark soliton, which produces destructive interference in second-order autocorrelation measurements.
\begin{figure*}
    \centering
    \includegraphics[width=0.8\textwidth]{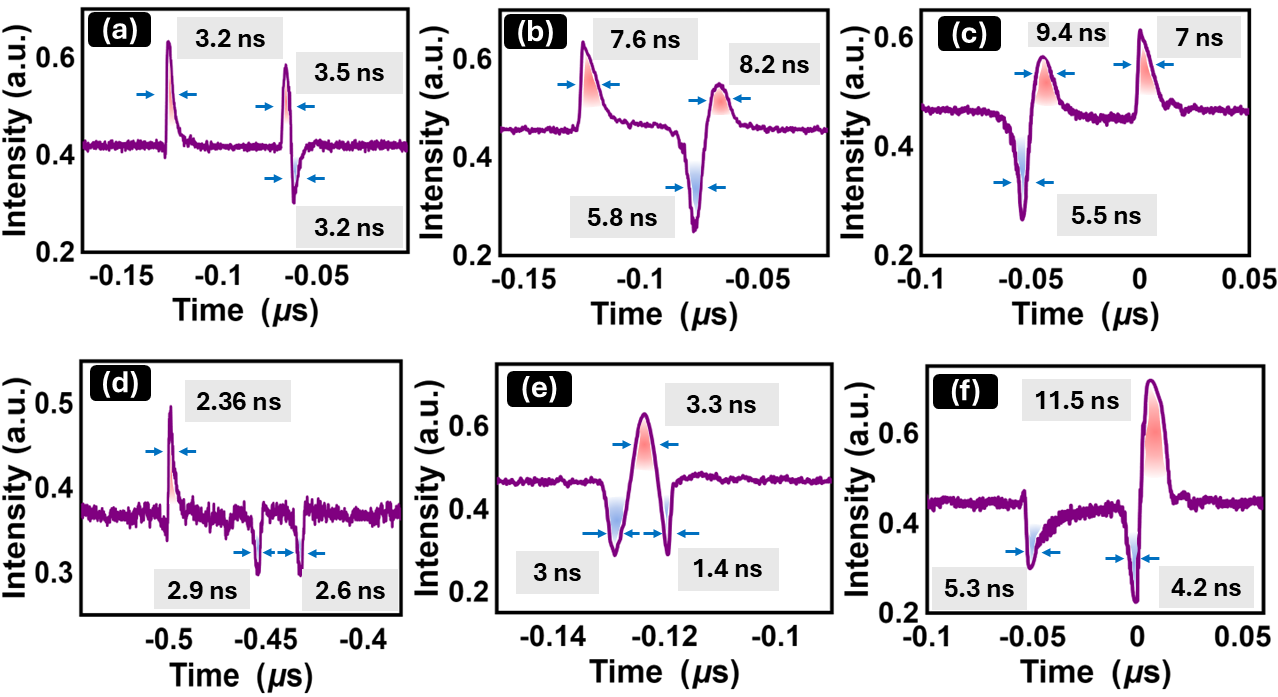} 
    \caption{Single-pulse temporal profiles of all six three-component 
    hybrid soliton states generated in the NPR-based EDFL: 
    (a)~bright-bright-dark~(BBD); 
    (b)~bright-dark-bright~(BDB); 
    (c)~bright-dark-dark~(BDD); 
    (d)~dark-bright-dark~(DBD); 
    (e)~dark-bright-bright~(DBB); 
    (f)~dark-dark-bright~(DDB). 
    Bright and dark soliton components are highlighted in red and 
    blue, respectively.}
    \label{5} 
\end{figure*}

Continuous rotation of the PCs at fixed pump power enables smooth and reproducible 
modulation of the relative intensities and temporal separations of the two soliton 
components. This tunability originates from PC-induced modifications to the NPR 
transmission window, which reshape the intensity- and wavelength-dependent loss 
experienced by each polarization eigenstate independently. Complete decoupling of the bright and dark components into independently propagating entities is achievable exclusively through polarization adjustment — without any pump power variation — providing direct evidence that the DB coupling is governed by XPM-mediated trapping rather than direct temporal overlap. The linearly polarized character of both components further establishes the DB pair as a group-velocity-locked vector soliton state.

Reversing the PC rotation direction at constant pump power induces a reproducible 
DB-to-BD transition~[Fig.~\ref{3}(a)], reflecting the asymmetry of the NPR 
transmission curve with respect to the polarization ellipse orientation. This reversal 
effectively exchanges the roles of the fast and slow polarization eigenstates, inverting the relative temporal ordering of the bright and dark components. An offset-DB configuration~[Fig.~\ref{3}(b)], in which the two components propagate with a 
well-defined temporal separation, reveals resolvable dual emission peaks at slightly 
offset center wavelengths — consistent with the group-velocity-locking mechanism, 
whereby XPM phase-matches two spectral components at different carrier frequencies 
to equalize their group velocities and sustain stable co-propagation. Rotating the PC back toward the original orientation causes the separated components to converge and 
merge into a tightly coupled DB pair, vividly demonstrating polarization-controlled 
XPM-mediated soliton trapping and release within the NPR cavity.
\subsection{Higher-Order Three-Component Hybrid Soliton Complexes}
As the pump power was progressively increased to the range of 240-280~mW, the systematic rotation of the PCs reproducibly unlocked a hierarchy of three-component hybrid soliton complexes. In this pump power regime, in which the intracavity gain substantially exceeds the threshold required for three-component soliton formation, and the NPR transmission window becomes sufficiently nonlinear to support simultaneous multi-eigenstate excitation, we observed all accessible permutations of bright and dark soliton ordering. These higher-order states --- bright-bright-dark~(BBD), 
bright-dark-bright~(BDB), bright-dark-dark~(BDD), 
dark-bright-dark~(DBD), dark-bright-bright~(DBB), and 
dark-dark-bright~(DDB) --- represent a comprehensive and 
systematically accessible family of multi-component dissipative 
solitons, the single-pulse temporal profiles of which are presented in 
Fig.~\ref{5} (a)-(f). To the best of our knowledge, this 
constitutes the first simultaneous experimental realization of all possible six 
permutations of the three-component bright-dark soliton family within 
a single laser cavity, and moreover, the first demonstration of any such 
states via an artificial saturable absorber mechanism. The ability to navigate this complete family by PC and pumped power adjustment, without changing cavity geometry, emphasizes the exceptional reconfigurability of NPR-based 
mode-locking as compared to the fixed-response saturable absorbers 
employed in all prior reports of multi-component hybrid 
states~\cite{Zhao2020, Wang2022, Shi2024, Wu2021, Ma2020}.\\

The observed three-component states exhibit rich internal structure governed by XPM and polarization-resolved gain dynamics. The BDB configuration, shown in Fig.~\ref{bdb_f}(a) along with its pulse widths extracted from full width at half maximum (FWHM) measurements, represents the most extensively studied three-component hybrid state in prior literature~\cite{Wu2021, Wang2022, Ma2020}. The dark soliton is sandwiched temporally between the two bright components, which are orthogonally polarized with respect to the dark component. This configuration is physically interpreted as a group-velocity-locked vector soliton~\cite{Wu2021}: the orthogonal components on the slow and fast eigenstates are phase-matched by XPM such that their respective group velocities are equalized, enabling stable co-propagation over many round trips without relative temporal drift. The notably wider temporal profile of BP2~(8.2~ns) relative to BP1~(7.6~ns) originates from the odd-symmetric phase structure of the intervening dark soliton: BP1 resides on the leading edge of the dark soliton phase profile and experiences a negative XPM phase gradient that spectrally compresses its profile, while BP2 resides on the trailing edge and experiences a positive phase gradient that induces spectral broadening and a correspondingly wider temporal extent. The M-shaped dual-peaked optical spectrum accompanying 
this state, as illustrated in Fig.~\ref{bdb_f}(c), with spectral components at 1567.64~nm and 1568.36~nm, is consistent with the group-velocity locking condition derived from the coupled nonlinear Schrödinger equation (CNLSE), which requires a finite wavelength separation 
between the two polarization eigenstates to compensate their 
differential group delay~\cite{Christodoulides1988}.


\begin{table}[htbp]
\centering
\caption{Measured FWHM pulse widths (ns) of three-component hybrid soliton states.}
\label{tab:three_component}
\small
\begin{tabular}{lcc}
\hline
\textbf{State} & \textbf{Widths (ns)} & \textbf{Ref.} \\
\hline
\multicolumn{3}{l}{\textit{Group I: Bright-dominant}} \\
BBD & 3.2, 3.5, 3.2 & This work \\
BDB & 7.6, 5.8, 8.2 & \cite{Wu2021,Wang2022} \\
DBB & 5.5, 9.4, 7.0 & \cite{Zhao2020} \\
\hline
\multicolumn{3}{l}{\textit{Group II: Dark-dominant}} \\
BDD & 2.36, 2.9, 2.6 & This work \\
DBD & 3.0, 3.3, 1.4 & This work \\
DDB & 5.3, 11.5, 4.2 & \cite{Shi2024} \\
\hline
\end{tabular}
\end{table}
The six states naturally partition into two groups based on 
their component composition. The first group --- BBD, BDB, 
and DBB --- is bright-dominant~(2B:1D), comprising two bright 
and one dark component. Since bright solitons are the 
energetically favored entities in the anomalous dispersion 
regime, sustained by the SPM-GVD balance, these states are 
relatively accessible and stable. In the BBD 
state~[Fig.~\ref{5}(a), widths: 3.2, 3.5, 
3.2~ns], two bright solitons co-propagate on the fast eigenstate 
with a trailing dark soliton on the slow eigenstate; the slight 
BP2 broadening~(3.5 vs 3.2~ns) encodes the additional phase 
load from the adjacent dark component. The BDB 
state~[Fig.~\ref{5}(b)], as we already discussed in detail.
In the DBB state~[Fig.~\ref{5}(c), widths: 
5.5, 9.4, 7.0~ns], the progressive BP narrowing from 9.4 to 
7.0~ns reflects the declining gain availability after partial 
saturation by the leading bright component; this state was 
previously reported~\cite{Zhao2020}, 
and the present NPR-based realization achieves a markedly 
superior SNR of 67.06~dB versus 35~dB therein.
 
The second group --- BDD, DBD, and DDB --- is dark-dominant~(1B:2D), 
comprising one bright and two dark components. Supporting two 
dark solitons simultaneously is physically more demanding, as 
both components compete for the same CW background field 
necessary to sustain a dark soliton notch, requiring more 
precise PC tuning and higher intracavity energy. In the BDD 
state~[Fig.~\ref{5}(d), widths: 2.36, 2.9, 
2.6~ns], a single bright soliton drives two successive dark 
components via cascaded phase modulation, with the progressive 
narrowing from DS1 to DS2~(2.9 to 2.6~ns) reflecting the 
diminishing CW background available at each successive 
nucleation step. The DBD state~[Fig.~\ref{5}(e), 
widths: 3.0, 3.3, 1.4~ns] is the structural inverse of BDB, 
with the pronounced dark width asymmetry~(3.0 vs 1.4~ns) 
directly encoding the directional asymmetry of the NPR 
transmission curve across the polarization ellipse orientation. 
In the DDB state~[Fig.~\ref{5}(f), widths: 
5.3, 11.5, 4.2~ns], recently reported using 
germanene~\cite{Shi2024}, the DS1-DS2 width 
disparity~(5.3 vs 11.5~ns) constitutes a quantifiable 
experimental signature of the range-dependent inter-component 
phase coupling: the dark component directly adjacent to the 
bright soliton experiences a more localized and intense phase 
gradient, yielding a significantly wider notch.
 
It is also worth noting that the two groups are related by 
simultaneous time-reversal and bright-dark exchange symmetry 
of the CNLSE: BBD~$\leftrightarrow$~DDB, 
BDB~$\leftrightarrow$~DBD, and DBB~$\leftrightarrow$~BDD. 
The experimental accessibility of both members of each mirror 
pair confirms that the NPR transmission curve symmetrically 
explores both sides of this fundamental symmetry of the 
coupled nonlinear wave equations. Of the six states, BBD, 
BDD, and DBD have no prior experimental precedent in any 
laser platform, while BDB, DBB, and DDB are demonstrated 
here for the first time via an artificial saturable absorber 
with substantially improved stability over material-SA-based 
reports~\cite{Wu2021, Wang2022, Zhao2020, Shi2024}. The 
complete accessibility of all six permutations through PC 
rotation alone directly validates the theoretical prediction 
of a continuously connected family of three-component CNLSE 
soliton solutions~\cite{Radhakrishnan1997} and establishes 
NPR mode-locking as the most versatile platform yet 
demonstrated for systematic exploration of multi-component 
hybrid soliton physics.

\subsection{Higher-Order Multi-Component Hybrid Soliton Complexes: 
DDBB, DBDB, and BDBDB States}

\begin{figure*}
    \centering
    \includegraphics[width=.8\textwidth]{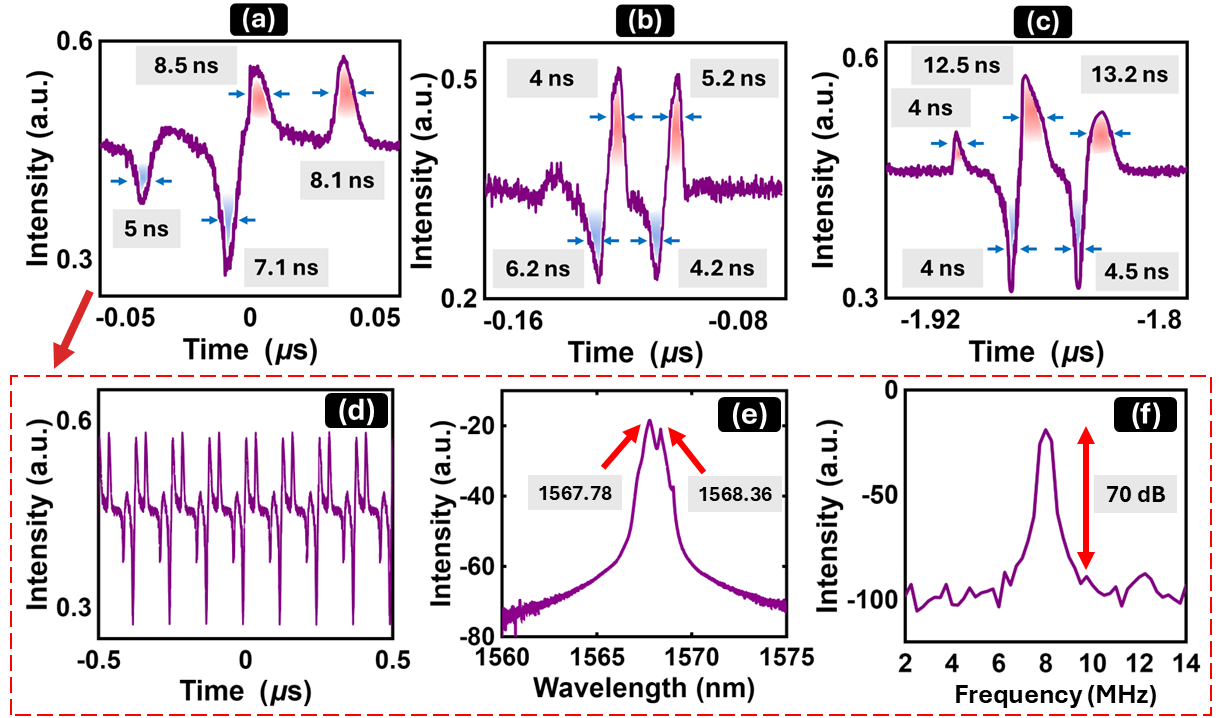} 
    \caption{Experimental characterization of four- and five-component 
    hybrid soliton states in the NPR-based EDFL. 
    Single-pulse temporal profiles of: 
    (a)~dark-dark-bright-bright~(DDBB), (b)~dark-bright-dark-bright~(DBDB), and  (c)~bright-dark-bright-dark-bright~(BDBDB) soliton along with 
    component widths. Red and blue shading denote 
    bright and dark soliton components, respectively. 
    Characterization of the DDBB state: 
    (d)~pulse train at fundamental repetition rate of 8.015~MHz; 
    (e)~optical emission spectrum exhibiting M-shaped dual-peaked 
    profile with dominant wavelengths at 1567.78~nm and 1568.36~nm; 
    (f)~RF spectrum at the fundamental repetition frequency with 
    SNR of 70~dB, confirming highly stable 
    mode-locked operation.}
    \label{6} 
\end{figure*}

Upon further optimization of the intracavity polarization state and pump power, the NPR-based mode-locked cavity demonstrated a remarkable capacity to support soliton architectures of even greater complexity than the three-component states described in the preceding section. By systematically rotating the PCs beyond 
the regimes that stabilize the three-component family, three unprecedented higher-order hybrid configurations were reproducibly accessed: the four-component dark-dark-bright-bright~(DDBB) and dark-bright-dark-bright~(DBDB) solitons, and the five-component bright-dark-bright-dark-bright~(BDBDB) soliton. The single-pulse 
temporal profiles of all three states are presented in Fig.~\ref{6}(a)-(c). The comprehensive characterization of the DDBB state, including pulse train, optical 
spectrum, and RF spectrum is presented in Fig.~\ref{6}(d)-(f). To the best of our 
knowledge, none of these states has been previously reported in any fiber laser system employing either material-based or artificial saturable absorbers, and their simultaneous realization within a single NPR-based erbium-doped fiber laser cavity constitutes a significant advance in the experimental taxonomy of dissipative multi-component solitons.

The complex DDBB state [Fig.~\ref{6}(a)] represents a higher-order extension of hybrid vector soliton dynamics, arising from the cooperative interplay of XPM, weak birefringence, and NPR-induced polarization mixing among multiple dark and bright components in the orthogonal eigenstates. These structures are subsequently phase-matched and mutually locked through inter-eigenstate XPM, which enforces group-velocity synchronization and stabilizes the composite temporal arrangement. The observed variation in component widths reflects the non-uniform XPM environment, where features experiencing stronger cumulative phase modulation broaden, while others remain comparatively narrow, providing a direct signature of cascaded nonlinear coupling within the cavity.
The stability and coherence of this state are confirmed by its uniform pulse train and high signal-to-noise ratio as illustrated in Fig.~\ref{6}(d)-(f), indicating that it constitutes a single dissipative eigenstate rather than a superposition of independent pulses. The characteristic dual-peaked optical spectrum further evidences inter-eigenstate group-velocity locking, with the spectral separation encoding the birefringence-induced delay compensated by XPM. Notably, the symmetric ordering of dark and bright features contributes to reduced polarization imbalance, enhancing robustness against perturbations. More broadly, the existence of this state highlights the ability of NPR-based cavities to sustain tightly coupled, higher-order soliton structures through continuous tuning of birefringence and nonlinear transmission, enabling access to regimes beyond those achievable with fixed-response saturable absorbers.

\begin{table}[htbp]
\centering
\caption{Comparative summary of higher-order hybrid soliton states 
observed in the present NPR-based EDFL, with comparison to prior 
experimental reports. N/A: not previously reported.}
\label{tab:higher_order_comparison}
\begin{tabular}{lccccc}
\hline
\textbf{State} & \textbf{Components}& \textbf{Widths (ns)}  & \textbf{Prior report} \\
\hline
DDBB & 4 & 5.0,~7.1,~8.5,~8.1 & N/A \\
DBDB & 4 & 4.0,~6.2,~5.2,~4.2 & N/A \\
BDBDB & 5 & 4.0,~12.5,~4.0,~13.2,~4.5 & N/A \\

\hline
\end{tabular}
\end{table}
\section{Numerical simulations and physical origin of dark-bright pulse complexes}
To understand the formation mechanism and stability of these complex dark-bright states, we perform numerical simulations that closely replicate the erbium-doped fiber ring cavity used in the experiment. The dynamics are modeled using the coupled CGLE~\cite{PhysRevA.81.053819} for the representative case of a DBD pulsed regime. The evolution of the orthogonally polarized field components $u(z,t)$ and $v(z,t)$ is governed by the coupled equations

\small
\begin{align}
\nonumber
\frac{\partial u}{\partial z} &= i\beta u - \delta \frac{\partial u}{\partial t}
- \frac{i\beta_2}{2}\frac{\partial^2 u}{\partial t^2} + \frac{g}{\Omega_g^2}\frac{\partial^2 u}{\partial t^2} +\\ &\frac{g}{2}u
+ i\gamma\left(|u|^2 + \frac{2}{3}|v|^2\right)u
+ \frac{i\gamma}{3}v^2 u^*, \\
\nonumber
\frac{\partial v}{\partial z} &= -i\beta v + \delta \frac{\partial v}{\partial t}
- \frac{i\beta_2}{2}\frac{\partial^2 v}{\partial t^2} + \frac{g}{\Omega_g^2}\frac{\partial^2 v}{\partial t^2} +\\
&\frac{g}{2}v
+ i\gamma\left(|v|^2 + \frac{2}{3}|u|^2\right)v
+ \frac{i\gamma}{3}u^2 v^* ,
\end{align}
\normalsize
where $\beta$ denotes the phase birefringence, $\delta$ represents the inverse group-velocity mismatch between the two polarization modes, $\beta_2$ is the group-velocity dispersion, and $\gamma$ is the Kerr nonlinear coefficient.  Here, the factor of $2/3$ in the XPM terms arises from the $\chi^{(3)}$ tensor symmetry of the isotropic silica medium. The self-phase modulation~(SPM) terms ($i\gamma|u|^2 u$) govern the formation of individual bright solitons in each polarization 
eigenstate through the balance with anomalous GVD, whereas the 
XPM terms ($i\gamma\frac{2}{3}|u|^2 u$) introduce an intensity-dependent refractive index modulation that couples the two polarization components nonlinearly.  Both the saturable gain and spectral filtering are included, modeled using a standard energy-dependent saturation form,
\begin{equation}
g = \frac{g_0}{1 + E/E_{\mathrm{sat}}},
\end{equation}
where $g_0$ is the small-signal gain coefficient, $E = \int (|u|^2 + |v|^2)\,dt$ is the intracavity pulse energy, and $E_{\mathrm{sat}}$ is the saturation energy. The finite gain bandwidth is incorporated through a second-order spectral filtering term. In passive fiber segments, the gain is set to zero and replaced by linear loss. The CGLE through each fiber segments are solved by employing the standard split-step Fourier method.

In contrast to a purely distributed model, the cavity is implemented using a lumped-element map to reflect realistic laser operation. The modeled cavity consists of a segment of erbium-doped fiber (EDF) of length $L_{\mathrm{EDF}} = 3\,\mathrm{m}$ and multiple single-mode fiber (SMF) sections with total length $L_{\mathrm{SMF}} = 22.5\,\mathrm{m}$, forming a dispersion-managed ring cavity. The group velocity dispersion (GVD) coefficients were set to $\beta_{2}^{\mathrm{EDF}} = 41\times10^{-27}\,\mathrm{s^2/m}$ (normal dispersion) and $\beta_{2}^{\mathrm{SMF}} = 21.7\times10^{-27}\,\mathrm{s^2/m}$ (anomalous dispersion), while nonlinear coefficients were calculated from $n_2 = 2.6\times10^{-20}\,\mathrm{m^2/W}$ using the effective mode areas of each fiber. Linear losses for the SMFs were taken as $\alpha = 2.3\times10^{-4}\,\mathrm{m^{-1}}$. The gain fiber was modeled with a saturable gain $g = G_0/(1+E/E_{\mathrm{sat}})$, where the small-signal gain coefficient was $G_0 = 0.12\,\mathrm{m^{-1}}$ and the saturation energy $E_{\mathrm{sat}} = 50\,\mathrm{pJ}$, together with a finite gain bandwidth corresponding to $\Delta\lambda_g = 50\,\mathrm{nm}$. The cavity output coupling ratio was fixed at $10\%$.

Nonlinear polarization rotation is modeled using a scalar intensity-dependent transmission function, as given below, providing an effective saturable absorber mechanism. 
\begin{align}
T(t) &= \frac{a_0}{1 + \dfrac{|u(t)|^2 + |v(t)|^2}{P_{\text{sat}}}} + n_s,
\end{align}
Where, $a_0$ is the modulation depth, $n_s$ represents the non-saturable losses and $P_{sat}$ is the saturation power. The values for these parameters used in the simulation are 0.8, 0.1 and 10 W, respectively. To excite a DBD structure, the initial condition is constructed as a composite field consisting of a central bright pulse embedded within a double-dark background,
\begin{align}
u(t) &= A_{\mathrm{cw}} + A_b \,\mathrm{sech}\left(\frac{t}{t_0}\right), \\
v(t) &= A_d \,\tanh\left(\frac{t}{t_0}\right)\tanh\left(\frac{+t_s}{t_0}\right),
\end{align}
where $t_0$ defines the characteristic pulse width and $t_s$ determines the separation between the two dark dips. This configuration corresponds to two domain walls located symmetrically about the origin, forming a localized region in which a bright pulse can be trapped.

\begin{figure*}
    \centering
    \includegraphics[width=\textwidth]{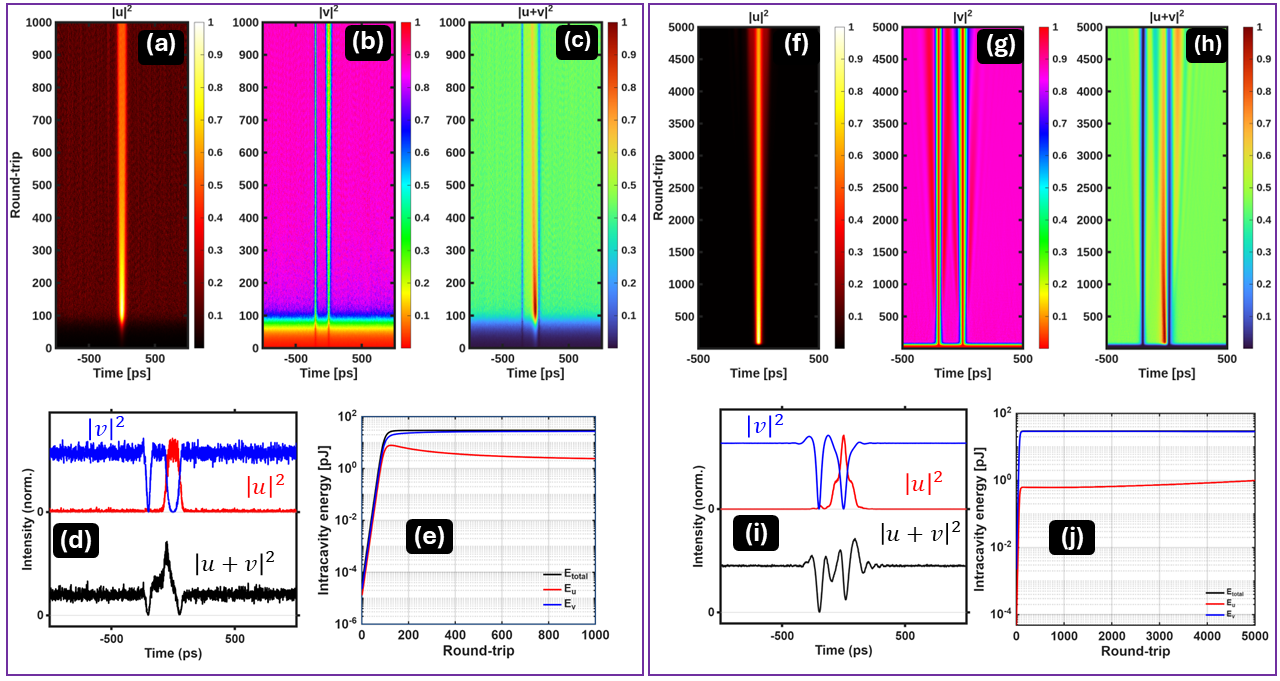} 
    \caption{Dynamics of the dark-bright-dark (DBD) pulse for an unevenly seeded initial condition (a)-(e). Panels (a) and (b) show the temporal evolution of the orthogonally polarized components, while (c) shows the evolution of the combined intensity. Panel (d) presents the final temporal profile after 1000 round-trips (RTs), and (e) shows the intracavity energy per round-trip. Panels (f)-(j) display the corresponding dynamics for a symmetrically seeded DBD pulse.}
    \label{simu} 
\end{figure*}

The temporal field evolution every roundtrip reveals the formation of a robust DBD vector state under different initial conditions. Fig.~\ref{simu} (a)–(e) shows the case where a weak dark seed is imposed on one polarization component, while the orthogonal component supports a bright pulse, forming an uneven seed. Despite the asymmetry of the initial condition, the system rapidly converges toward a stationary DBD configuration within the first few hundred round-trips, as corroborated by the energy convergence [Fig.~\ref{simu}(e)]. A similar evolution is observed in  Fig.~\ref{simu}(f)–(j), where both dark and bright components are initialized with equal amplitudes (symmetrically distributed amplitudes). In this case, although the transient dynamics differ slightly, the system evolves toward the  DBD attractor, and after approximately 1000 round-trips, exhibits a slight temporal broadening of the bright component, accompanied by minor changes in intracavity energy [Fig.~\ref{simu}(j)]. The final temporal pulse reveals a complex DBD state, indicating the presence of residual dispersion that is not fully compensated by the nonlinear and dissipative effects. Consequently, the DBD structure should be regarded as quasi-stationary: its overall shape and relative configuration are preserved over many round-trips, while its width evolves slowly due to imperfect dispersion balance.

Moreover, the formation and sustenance of the three-component soliton state are primarily governed by the simultaneous operation of two distinct XPM-mediated processes within a single cavity round trip: \textit{(i)} XPM-induced bright soliton trapping, whereby the intense bright component induces an effective potential well through the $2/3$ cross-coupling term that confines and stabilizes the orthogonally polarized field; and \textit{(ii)} XPM-induced dark soliton nucleation, whereby the same cross-coupling generates a $\pi$-phase-shifted 
notch in the CW background of the orthogonal component, seeding and sustaining a dark soliton or multiple dark solitons. In addition, the weak birefringence ensures sufficient group-velocity mismatch between the polarization modes, enabling sustained interaction while preserving their vector nature~\cite{ZHANG2026115152}. In the resulting multi-component state, the position of the bright and dark components is determined by the relative phase, group velocity, and amplitude of the two polarization eigenstates, all of which are continuously and independently tunable through the intracavity PCs. This is the fundamental reason why NPR, unlike fixed-response material-based saturable absorbers, can access all six permutations of the three-component family: the PC rotation continuously reconfigures the NPR transmission ellipse, modifying the intensity-dependent loss experienced by each polarization component and thereby selecting the nonlinear eigenstate that the cavity self-consistently converges to.

The formation of dark soliton components within the anomalous dispersion cavity warrants particular physical discussion. In the anomalous dispersion regime, the conventional expectation is that bright solitons are stabilized by the SPM-GVD balance, while dark solitons are associated with normal dispersion. However, in the presence of birefringence and XPM, a dark soliton on one polarization eigenstate can be sustained even in anomalous net dispersion, provided 
that the XPM-induced phase modulation from the orthogonally polarized bright component supplies the necessary nonlinear phase gradient to support the dark notch against dispersive broadening. This XPM-sustained dark soliton mechanism, first theoretically analyzed by Christodoulides and Joseph~\cite{Christodoulides1988} and subsequently confirmed experimentally in two-component systems~\cite{Zhao2010, Ning2012}, is extended in the present work 
to the three-component regime, where two such XPM interactions 
operate simultaneously and competitively within the same round trip.
Moreover, the emergence of four- and five-component soliton complexes from the cavity can also be attributed to the coupling between multi-soliton states of the orthogonal polarization components. Interestingly, the CGLE governing the two orthogonal polarization components of the intracavity field admits multi-soliton solutions on each polarization eigenstate simultaneously. The theoretical basis for the existence of such higher-order multi-component coupled soliton states was established by Radhakrishnan \textit{et~al.}~\cite{Radhakrishnan1997}, who 
demonstrated through exact solutions of the CNLSEs that $N$-component soliton states exist for arbitrary $N$, provided that the group-velocity matching condition and inter-component phase relations are simultaneously satisfied. The present experimental results provide the first laboratory verification of this prediction beyond the three-component level.

The numerical results demonstrate that stable DB pulse complexes can be sustained in passively mode-locked fiber laser cavities. These states arise from the interplay of nonlinear coupling, gain saturation, and birefringence-controlled polarization dynamics, extending the family of vector soliton solutions beyond conventional dark-bright and bright-dark-bright configurations.

\section{Summary}
In this study, the dynamics of DB soliton pairs and higher-order soliton structures were extensively investigated in an NPR mode-locked fiber laser. The experimental findings demonstrated that by systematically tuning the pump power and PC, various soliton states, such as DDB, DBD, DDBB, BDBDB, etc., could be generated and manipulated. The ability to control these soliton configurations highlights the flexibility of mode-locked fiber lasers in producing complex multi-soliton interactions.

The formation of these soliton states was found to be driven by a combination of key physical effects, including dispersion, nonlinearity, birefringence-induced polarization effects, and XPM. The self-consistent nature of mode-locking ensured that the repetition rate remained stable across various pump power levels, further validating the robustness of the generated soliton structures. The optical and RF spectra confirmed the multi-wavelength emission characteristics of the solitons, with stable and high SNR, indicating strong mode-locking stability.

A significant aspect of this work was the demonstration of the tunability of soliton interactions. The experimental results revealed that dark and bright solitons could propagate independently along the fiber axis and could be merged or separated by adjusting the PC settings. This tunability suggests that polarization-dependent nonlinear effects, particularly XPM, play a critical role in the interaction and stabilization of these solitons. Additionally, the emergence of asymmetric spectral profiles, including Kelly sidebands, further confirmed the complex interplay of gain, dispersion, and nonlinearity in soliton formation.

Furthermore, numerical simulations confirmed XPM-driven cross-polarization coupling between multi-soliton states, leading to complex DB structures. Overall, this work provides valuable insights into the complex nonlinear dynamics of soliton interactions in fiber lasers. The results contribute to the broader understanding of soliton formation mechanisms, which can be applied to the development of high-performance ultrafast photonic systems. Future research will focus on further optimizing the laser cavity parameters, exploring additional nonlinear optical effects, and extending the study to different saturable absorber materials to achieve even greater control over soliton dynamics and enhance the capabilities of mode-locked fiber lasers for advanced optical applications.
\section{Conclusion}
In this work, we experimentally demonstrate a wide range of DB soliton complexes, including higher-order configurations, in an NPR mode-locked fiber laser through controlled tuning of pump power and polarization. The observed states arise from the interplay of dispersion, Kerr nonlinearity, birefringence, XPM, and NPR-induced dissipative effects, enabling flexible manipulation and stabilization of complex multi-component soliton structures with high signal-to-noise ratios.

These experimental findings are further supported by numerical simulations based on the CGLE, which reproduce the formation and stability of representative states such as the DBD structure and confirm their nature as dissipative attractors of the cavity dynamics. The combined experimental and numerical results establish NPR-based fiber lasers as a highly versatile platform for exploring multi-component vector soliton physics, with potential applications in high-energy ultrafast photonics and advanced nonlinear optical systems.

\section{Acknowledgements}
The authors acknowledge CEFIPRA/IFCPAR (IFC/7148/2023) and UKIERI-SPARC (3673) for financial support through research projects. Additionally, KN thanks the Anusandhan National Research Foundation~(ANRF) for support through the Core Research Grant (CRG/2023/008068). Subrata Manna (Ref- 2003349) and Amala Jose ~ (Ref- 2002215) gratefully acknowledge PMRF for the financial support through the PMRF fellowships.

\bibliographystyle{IEEEtran}
\bibliography{Refs}
\end{document}